\documentclass[AMA,STIX1COL]{WileyNJD-v2}
\usepackage{moreverb}
\usepackage{subcaption}
\usepackage[linesnumbered,algoruled,boxed,lined]{algorithm2e}
\SetKw{kwInit}{Initialization:}

\newcommand\BibTeX{{\rmfamily B\kern-.05em \textsc{i\kern-.025em b}\kern-.08em
T\kern-.1667em\lower.7ex\hbox{E}\kern-.125emX}}

\articletype{Research Article}%

\received{<day> <Month>, <year>}
\revised{<day> <Month>, <year>}
\accepted{<day> <Month>, <year>}

\begin{document}

\title{GeoGraphViz: Geographically Constrained 3D Force-Directed Graph for Knowledge Graph Visualization}

\author[1,2]{Sizhe Wang}

\author[1]{Wenwen Li*}

\author[1]{Zhining Gu}

\authormark{Wang \textsc{et al}}

\address[1]{\orgdiv{School of Geographical Sciences and Urban Planning}, \orgname{Arizona State University}, \orgaddress{\state{Arizona}, \country{USA}}}

\address[2]{\orgdiv{School of Computing and Augmented Intelligence}, \orgname{Arizona State University}, \orgaddress{\state{Arizona}, \country{USA}}}

\corres{*Wenwen Li. School of Geographical Sciences and Urban Planning, Arizona State University, Tempe, AZ, ASU. \\
\email{wenwen@asu.edu}}


\abstract[Abstract]{Knowledge graphs are a key technique for linking and integrating cross-domain data, concepts, tools, and knowledge to enable data-driven analytics. As much of the world’s data have become massive in size, visualizing graph entities and their interrelationships intuitively and interactively has become a crucial task for ingesting and better utilizing graph content to support semantic reasoning, discovering hidden knowledge discovering, and better scientific understanding of geophysical and social phenomena. Despite the fact that many such phenomena (e.g., disasters) have clear spatial footprints and geographical properties, their location information is considered only as a textual label in existing graph visualization tools, limiting their capability to reveal the geospatial distribution patterns of the graph nodes. In addition, most graph visualization techniques rely on 2D graph visualization, which constraints the dimensions of information that can be presented and lacks support for graph structure examination from multiple angles. To tackle the above challenges, we developed a novel 3D map-based graph visualization algorithm to enable interactive exploration of graph content and patterns in a spatially explicit manner. The algorithm extends a 3D force directed graph by integrating a web map, an additional geolocational force, and a force balancing variable that allows for the dynamic adjustment of the 3D graph structure and layout. This mechanism helps create a balanced graph view between the semantic forces among the graph nodes and the attractive force from a geolocation to a graph node. Our solution offers a new perspective in visualizing and understanding spatial entities and events in a knowledge graph.}

\keywords{knowledge graph; force-directed graph; visualization; graph patterns; linked data; location-aware}

\maketitle


\section{Introduction}
Knowledge graphs are a machine-parsable graph technology that interlinks world entities, both physical and digital, and provides formal descriptions of each entity to enhance information consumption, support query answering, and semantic inference and reasoning for deriving new knowledge \cite{li2020geoai, li2019ontology}. As the data volume in geospatial and other domains continue to hit exponential growth, organizations and data holders have increasingly used knowledge graphs to organize massive amounts of domain data. The IT giants, such as Google, have created a knowledge graph by integrating information from a variety of sources to support better search and information retrieval. Statistics have shown that by May 2020, Google’s knowledge graph has grown to contain 5 billion entities and over 500 billion facts, and it has supported over 1/3 of 100 billion queries made to Google each month.

In the geospatial domain, progress in building spatial data infrastructures, such as the European Union’s INSPIRE (Infrastructure for Spatial Information in the European Community) portal, has been in creating data and metadata schemas to annotate data in the broad Earth and environmental science domains \cite{tschirner2011semantic}. In the US, the National Science Foundation (NSF) has started a new program called Convergence Accelerator. One of its aims is to develop cutting-edge knowledge graph technologies for linking cross-domain data for building an open knowledge network that fosters convergence research. A key research advancement in spatial sciences is the creation of KnowWhereGraph \cite{janowicz2022know}, a knowledge graph that connects environmental datasets related to natural disasters, agriculture, and soil properties to understand the environment impacts on the society. Semantic enrichment services are provided to support semantic reasoning and question answering on top of its data store of over 12 billion facts \cite{liu2022knowledge}.

As the domain data in various knowledge graphs proliferate in the research community, challenges also arise for non-developers as well as knowledge engineers to access and understand massive graph-ready datasets \cite{wang2019capturing} \cite{li2022performance} \cite{li2022geoai}. In particular, it has become increasingly challenging to enable open access to the right graph data in the right form for the right end users. As human brains are especially good at visual inspection, providing an intuitive presentation of the knowledge graphs are critical for the understanding of the graph, its entities, and interconnections among them. Knowledge graph visualization, which enables interactive exploration and information filtering of graph data and structure, has become an important means to facilitate data consumption and sense making of big, linked data.

Popular graph visualization tools, such as those provided as part of a triple store platform (e.g., GraphDB \cite{graphDBnew} or AllegroGraph \cite{aasman2006allegro}), mainly support 2D graph visualization. In the graph, nodes are used to represent the entities, and edges represent interrelationships between the entities. To allow real-time rendering, especially in handling large graph data, nodes expansion or clustering functions are provided such that the graph starts from a summarized view and can be gradually expanded according to end user interest. One major limitation in such solutions is the lack of geographical reference for the graph data, especially for data that are geospatial, such as hurricane tracks that have originated in the North Atlantic Ocean and moved westward to hit the US coastal communities, or an interruption of a food supply chain caused by a wildfire event that has occurred near the farmlands in Southern California. In existing visualization tools—even if the location information, which could be a point, line, or polygon—is encoded in the graph, they are treated as regular text and there are very few tools that support the visualization of this special graph information in a virtual geographical space, such as a map. 

In comparison, traditional GISs (Geographic Information Systems) provide powerful solutions for overlaying multiple geospatial data on a base map, and they often adopt flow maps to support graph data visualization \cite{hu2014linked, kim2012developing}. In such systems, when presenting interrelationships among entities, each entity must be pinned to a fixed geographic location on the map, regardless of whether an entity possesses multiple geographical properties or is a moving target (e.g., hurricane events). The relationships between the entities are visualized through straight or curved lines between two locations on a map. Although these map-based graph visualizations offer a geographical reference, the graph patterns, such as local clusters, are significantly altered as strengths of the edges connecting the nodes are not considered. This limitation hinders the dynamic exploration of graph structure and patterns. 

To overcome this limitation, this paper presents a new “Graph Above the Map” solution to enable interactive graph exploration from both the graph and map perspectives. A geographically constrained 3D force-directed graph visualization algorithm is developed to dynamically render the graph layout by considering the interrelations among the graph nodes and their geolocations. This way, both the location property and graph properties can be visualized in a coherent web interface, supporting knowledge discovery within a geospatial context. The rest of the paper is organized as follows: Section 2 reviews existing techniques for graph visualization, Section 3 introduces our proposed algorithm in detail, Section 4 describes the data and experiments to quantitatively measure the graph and geographical presentation of linked data, Section 5 introduces the GeoGraphViz user interface, and Section 6 concludes the paper with a discussion of future research directions. 
\section{Literature Review}
\subsection{Map-based visualization}
Flow maps are a prominent approach that uses map-based visualization to display linkages and relationships between two entities. It is a combination of a 2D or 3D map with a flow diagram. For instance, the Sankey diagram \cite{riehmann2005interactive} is type of flow diagram to illustrate a flow trend from the start node to the end node. In the flow diagram, the arrow on the edges shows the flow direction and the width of edges connecting each pair of nodes indicates the intensity between the connections. The edges can also be assigned with different colors to show different types of connections. When the start and end nodes are associated with some locations and are placed on a map, the diagram becomes a flow map. Relying on distinctive flow styles and node symbolization, a flow map can exhibit common and unique properties of different locations \cite{hu2014linked}. It can also intuitively demonstrate the patterns of flow, be it local, regional, or global. Because of the intuitive manner for displaying flows and connections, flow map visualization has received widespread adoptions in several applications, such as mapping traffic patterns, migration routes, trade patterns, disease spread, and other geospatial data with linkage information \cite{boyandin2010using,guo2009flow,luo2014geo}. To further improve the visual effect, Yang et al. \cite{8440844} proposed a 3D flow map to incorporate a height dimension of the arrows to display additional data attributes that 2D flow maps are not capable of. Ardissono et al. \cite{ardissono2018map} developed a novel model using colored shapes and interactive functions to assist entity filtering based on selected categories to better support map-based visualization. Because of its ability in presenting connections between points, flow maps can be leveraged to support graph visualization, particularly when location is the most prominent property of a graph node, such as a city. However, because a flow map needs to pinpoint the graph nodes to a fixed location on the map, it suppresses the display of some graph patterns, such as clustering patterns of the graph nodes measured by the strengths of their interconnections rather than the geographic proximity among the nodes. The flow map has also shown limitation in visualizing the spatiotemporal evolution of dynamic entities. 

\subsection{Graph visualization methods}
Recently, several graph visualization tools have also been developed to visualize linked data in a knowledge graph \cite{gomez2018visualizing, ji2021survey}. WebVOWL is a web-based visualization tool for ontologies \cite{lohmann2014webvowl}. It can directly load ontologies (a logical representation of knowledge graph) in an RDF (Resource Description Framework) format to visualize the interrelationships among the graph nodes. Different line styles and node colors are applied to differentiate entities, their properties and relationships. Several semantic graph databases, such as GraphDB \cite{guting1994graphdb}, Neo4j \cite{webber2012programmatic}, and ArangoDB \cite{fernandes2018graph}, also provide visual graph interfaces along with data query support to allow visualization of graph data. Similar to WebVOWL, these graph databases also provide 2D visual graphs to display nodes and their connections; some provide additional statistical functions to show counts of nodes and relationships belonging to different categories while others support filtering operation to visualize a subgraph of interest. In ArangoDB, graph visualization is combined with timeline visualization for identifying events and relationships with a temporal pattern. These built-in visualization solutions are based on drawing information on a 2D visual graph, limiting the multi-dimensional exploration of the graph. 

Besides these built-in tools for knowledge graph visualization, customized tools that support different graph exploration needs have also been developed. For instance, Heim et al. \cite{heim2008graph, heim2009relfinder} developed RelFinder to support the identification and visualization of node relationships. This is achieved through highlighting a path connecting two nodes of interests in a 2D graph. He et al. \cite{he2019aloha} developed an interactive graph platform to support knowledge discovery on dietary supplements. The graph can also be dragged and panned, and a node’s visibility can be configured. Noel et al. \cite{noel2016cygraph} developed CyGraph, a unified graph-based system for network security monitoring and improvement. CyGraph provides multiple functions to allow visual interactions with the graph, including node and edge property filtering and color configuration. Most interestingly, it can also show a cluster view of graph nodes that share the same property. Similar works that allow display of clustering patterns can also be found in \cite{hu2021data}. All these graph solutions are based on 2D visualizations.

There are also works integrating map visualization and graph visualization for linked data exploration. Liu et al. \cite{liu2022ld} developed a web portal that uses multiple visualization methods, such as 2D graph visualization, streamgraph visualization, map visualization, and circular flow charts, to showcase patterns of research collaboration, popular research topics, and other information from publications. Regalia et al. \cite{regalia2018gnis} developed a linked data browser called Phuzzy.link to use hyperlinks to hop among connected graph nodes. If a graph node has geographic information, a map will be displayed on the side to enrich the visualization \cite{regalia2017phuzzy}. This is a way to visualize graph data in a non-graph form. A similar visualization portal was developed by Mai et al. \cite{mai2022narrative}. In addition to visualizing graph data in a non-graph form, the authors employ a narrative cartography to map geographic information with timestamps on the map to tell a story, such as one describing an expedition path. 

Recently, 3D graph visualization has become a trending technique. Open-source JavaScript (JS) libraries, such as ThreeJS \cite{dirksen2013learning} and Gephi \cite{bastian2009gephi}, have become popular tools to provide 3D rendering engines using WebGL (Web Graphic Language) to visualize and explore graph data in 3D on the Web. Powered by ThreeJS, a force-directed 3D graph drawing mechanism can be implemented to place nodes in a visually 3D space to avoid node cluttering problems. The 3D forced-directed graph supports multiple ways of graph interactions and visual effects. For instance, users can drag the graph nodes and place them in a different position on the screen; they can also rotate the entire graph to observe patterns from different viewpoints (e.g., to view the graph from the front, the top, and the back, as well as in any other 3D angle). When a user clicks on a node, the information about it pops up, and the clicked node will be enlarged or highlighted. The nodes and edges can also be colored according to the types of connections. Text, images, and customized geometries can be used to render the nodes. These 3D visualization libraries offer an efficient way to explore graph data interactively in a web-based environment.

Existing visualization tools or libraries provide visual displays for users to interact and explore interested data, relationships, and knowledge within a graph. To allow real-time rendering of large graph data, node folding/expansion or clustering functions are often provided so that the visual graph will start from a summarized view and can be gradually expanded according to end user interest in certain subgraphs. However, in almost all these tools, location information is rendered as text information (e.g., as an annotation indicating the placename). This limitation makes it difficult to examine the geospatial relationships among the graph nodes or exploring the spatial contexts of semantically relevant information. As a result, it is difficult to intuitively exhibit the knowledge connections between spatial information and semantic relationships among entities. 

To address the aforementioned limitations in both the more traditional map-based visualization (which lacks the flexibility in presenting semantic relationships of the graph nodes) and 2D graph visualization (which lacks the intuitive representation of location information and spatial context), we propose in a new “Graph Above the Map” solution, the GeoGraphViz. In this, 3D web-based graph visualization is enabled in combination with a 2D map view to allow a more coherent spatial-semantic visual presentation of knowledge graph data. In particular, a geographically constrained 3D force-directed graph visualization algorithm is developed. The algorithm is described in detail in the next section. 

\section{GeoGraphViz Visualization Algorithm} \label{sec_alg}

The intuition  behind our proposed algorithm is that current graph visualization strategies almost entirely visualize nodes and links based on their semantic relationships. The weights on the relationships are often ignored in existing solutions. As such, nodes are placed close to each other not because they are similar, but because there are linkages among them. Force-directed graphs are capable of simulating the graph layout by the strengths of connections among nodes. Therefore, they are very suitable for visualizing graphs which have quantitative weights (such as similarities) on the edges. However, the graph can only visualize based on one kind of force; the algorithm is not capable of visualizing a graph when multiple forces present, such as the semantic similarity \cite{li2012semantic} among graph nodes, and the geographical force that a node receives based on its own geolocation properties. However, we argue that the semantic connections among the nodes and the geographical distribution patterns of the nodes are both important for inspecting graph patterns, especially when the graph nodes are geospatially related (e.g., each node represents a geographical entity, such as a natural feature) \cite{li2016integrated}\cite{li2020automated}. Hence, in our paper, we are looking to “blend in” different amounts of semantic and geographical forces to demonstrate multi-faceted characteristics of graph patterns in a single view to enable a new kind of geospatial knowledge discovery. 

The GeoGraphViz visualization algorithm derives from and extends the force-directed graph placement algorithm \cite{fruchterman1991graph}. A force-directed graph simulates the graph layout with two forces: attractive and repulsive force among graph nodes. Attractive force is used to place connected nodes visually close to each other; it exists when there is an edge connecting two nodes. Repulsive force exists universally among all node pairs in the graph, regardless of whether a connection is presented. It prevents the nodes from becoming too close to each other in the visual graph. With a joint effect of attractive and repulsive forces, the position of the graph nodes is dynamically adjusted until the graph reaches a stable state, in which all forces are at equilibrium. The resultant graph layout is capable of presenting node clustering patterns based on their interconnectivity. It also maintains sufficient distance among the nodes to ensure graph readability and interpretability.

However, when the nodes contain location information, which is important to understand the spatial context of the graph patterns, this algorithm no longer fulfills the requirement. Our algorithm addresses this limitation by adding the third force, which we call the ``geo-force,'' to the original force-directed graph placement algorithm to support location-aware graph visualization. The geo-force is a type of attractive force from a specific geolocation to a graph node. The geolocation can be a point location unique to a geographic entity, such as a city; it could also be the location property of other named entities, such as an expert who is related to a location (e.g., through his/her affiliation). Integrating the three forces- the attractive force ($f^A$), repulsive force ($f^R$), and the geo-force ($f^G$)-the visualization algorithm can not only reveal the interrelationships among the graph nodes, but also the geographic distribution of the nodes. Mathematically, the three forces can be computed as follows:

\begin{align}
\label{eq_attrfoce}
    f^A(u,v)&=\frac{\left \|C(u) - C(v) \right\|^2}{k} \cdot \frac{C(u) - C(v)}{\left \|C(u) - C(v) \right\|} \notag\\
    &=\frac{\left\|C(u) - C(v) \right\|(C(u) - C(v))}{k}
\end{align}

\begin{align}
\label{eq_replforce}
    f^R(u, v)&= -\frac{k^2}{\left\|C(u) - C(v)\right\|}\cdot\frac{C(u)-C(v)}{\left\|C(u) - C(v)\right\|} \notag \\
    &= -\frac{k^2(C(u) - C(v))}{\left\|C(u) - C(v)\right\|^2}
\end{align}

\begin{align}
\label{eq_geoforce}
    f^G(u)&= K\cdot\frac{\left\|G(u) - C(u)\right\|^2}{k}\cdot\frac{G(u) - C(u)}{\left\|G(u) - C(u)\right\|} \notag \\
    &= K\cdot\frac{\left\|G(u) - C(u)\right\|(G(u) - C(u))}{k}
\end{align}

where $u$ and $v$ are two nodes between which the forces apply. $C(\cdot)$ indicates the coordinate vector of a graph node in a virtual 3D space in which the graph is placed. $G(\cdot)$ refers to the geographical coordinate vector (e.g., latitude and longitude) representing the geolocation property of a graph node. Note that, before calculation, the geographic coordinates need to be projected into the aforementioned virtual 3D space in which the graph layout is updated. $||\cdot||$ computes the magnitude of a vector. $f^A (u,v)$ and $f^R (u,v)$ respectively define the attractive and repulsive forces that $v$ receives from $u$.  $k (>0)$ is a parameter that balances between the attractive and repulsive forces. The larger the $k$ is, the smaller attractive force is, and thus the stronger repulsive force a node will receive, resulting in a less localized layout. Parameter $K (\geq 0)$ controls the relative strength of the geo-force compared with the two other forces. The larger the $K$ is, the closer a graph node will be placed toward its actual geolocation on a map.    

To simulate the final graph layout, the location of each node will be updated through an iterative process until the 3D graph reaches a stable state: a graph-level force equilibrium. Below, we provide the pseudocode of our proposed algorithm. The input of the algorithm includes the graph $G$, an initial system temperature $T$, and a cooling parameter $\alpha$. Here the system temperature and cooling factor are introduced to simulate the annealing process in which a solid is being heated up by a high temperature so that all particles of the solid can be transferred into the liquid state. This is followed by a cooling process that slowly lowers the temperature until all particles reach a low energy ground state \cite{li2014p}. These parameters are also often used in a heuristic-based optimization algorithm to apply more aggressive search/move (a high initial $T$) at the beginning of the process and a careful fine-tuning (smaller $\alpha$) at a later stage to find the near-optimal solution. Here in the algorithm, the system temperature $T$ determines the maximal distance $d_T$ that a node can move at each iteration. There are five main steps at each iteration. Steps 1-3 calculate three individual forces that act on each node from all the other nodes according to Equations \ref{eq_attrfoce}, \ref{eq_replforce}, and \ref{eq_geoforce}. The net force jointly determined by the three forces are saved in $F(v)$ as a vector. Step 4 updates a node’s position. The moving direction of the node is along the direction given by the net force $F(v)$ and its moving distance is co-decided by the strength of the net force $||F(v)||$ and a maximum moving distance $d_T$. The system temperature $T$ decreases (Step 5) as the process goes on so that, just as with the annealing process, the graph layout is updated fast at the beginning and slower later for fine-tuning to eventually reach a stable state, when an optimal or a near-optimal graph layout is found. 

\begin{algorithm}
  \caption{Geographically Constrained Force-Directed Graph Simulation}
  \KwIn{Graph $G = (V, E)$} 
  \KwIn{Initial temperature $T$}
  \KwIn{Cooling parameters $\alpha$}
  \KwOut{Graph $G$ with updated layout}
  \BlankLine
  \For{$i \leftarrow 1$ \KwTo $n\_iterations$}{
    \emph{// Step 0: Initialize the net force $F$ for each vertex}\\
    \ForEach{vertex $v \in V$}{
        \textbf{let} $F(v) \leftarrow 0$
    }
    
    \emph{// Step 1: Calculate accumulated attractive force}\\
    \ForEach{edge $e \in E$}{
    
      \textbf{let} $u$ and $v$ be the two vertices of $e$ and $u$, $v \in V$\\
      $F(v) \leftarrow F(v) + f^A(u, v)$\\
      $F(u) \leftarrow F(u) + f^A(v, u)$
    }
    
    \emph{// Step 2: Calculate accumulated repulsive force}\\
    \ForEach{vertex $v \in V$}{
        \ForEach{vertex $u \in V$}{
            \lIf{$u \neq v$}{$F(v) \leftarrow F(v) + f^R(u,v)$}
        }
    }
    \emph{// Step 3: Calculate the geo-force}\\
    \ForEach{vertex $v \in V$}{$F(v) \leftarrow F(v)+f^G(v) $}

    \emph{// Step 4: Update the coordinates of each vertex in the virtual 3D space}\\
    \textbf{let} $d_T \leftarrow T$\\
    \ForEach{vertex $v \in V$}{
    $C(v) \leftarrow C(v)+\frac{F(v)}{||F(v)||}\cdot\min(d_T, ||F(v)||)$
    }
    
    \emph{// Step 5: cooling function (temperature decay)}\\
    $T \leftarrow (1 - \alpha)\cdot T$\\
  }
\end{algorithm}

\section{Data and Experiments} \label{sec_exp}
\subsection{Graph Data}
\begin{figure}[h]
  \centering
  \includegraphics[width=\linewidth]{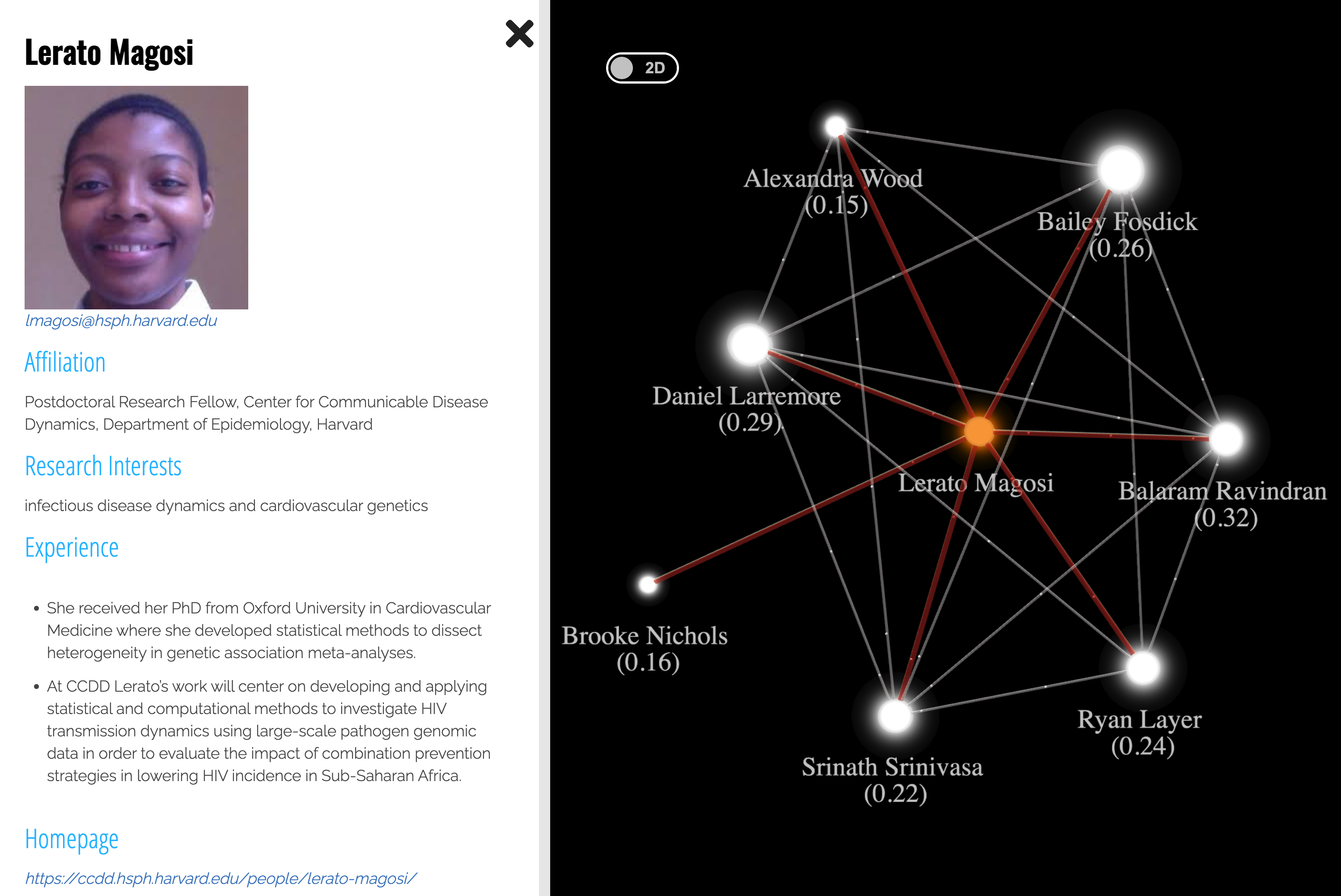}
  \caption{An example expert node in the expert graph.}
  \label{fig1}
\end{figure}

To test our proposed 3D geographically constrained force-directed graph rendering algorithm, we use expert network data from Direct Relief, a non-profile organization dedicated to providing disaster relief and humanitarian aid. This expert network contains 41 worldwide experts with infectious disease related expertise (e.g., COVID-19) that the Direct Relief staff works with to distribute medical and other supplies to help vulnerable and COVID-19 affected populations. Each expert has properties that include a name, research interest, affiliation and its geolocation, as well as a public research profile. The experts are linked through a semantic similarity measure of their research interests, generating a similarity graph. The similarity scores between two experts have a value range (0,1] and they are computed from a semantic analysis of the experts’ most representative publications. Figure \ref{fig1} shows the profile of an infectious disease expert (left) and her potential collaboration network with researchers sharing similar expertise (right).

\subsection{Force balancing and graph presentation} \label{subsec_exp_real}

To investigate the effect of the added geo-force to the visual presentation of the 3D graph, we conducted an experiment to measure the graph layout changes when the forces are assigned with different weights (parameter $K$ in equation \ref{eq_geoforce}, which is also known as the forcing balancing parameter). Here we propose two quantitative metrics to evaluate the degree to which the graph structure can be preserved and how well the spatial patterns can be revealed with an added geo-force. To measure the layout changes of a rendered graph, we adopt edge length variation (ELV), which provides a normalized measure of average edge length changes of a graph. This measure has been increasingly adopted in recent studies to determine the quality of a graph layout \cite{kwon2017would, haleem2019evaluating}. The metric $M_{ELV}$ is defined as follows:

\begin{align}
    M_{ELV} = \frac{l_v}{\sqrt{n_{E} - 1}}
\end{align}

with:

\begin{align}
   l_{v} = \sqrt{\sum\limits_{e \in E}\frac{(l_e - l_{\mu})^2}{n_E\cdot l^2_\mu}}
\end{align}

Where $E$ is the set of edges in the graph, $n_E$ is the number of edges, and $l_{\mu}$ is the mean length of all edges. Terms $\sqrt{n_E-1}$ and $l_{\mu}^2$ are added for normalization purposes. The intuition behind our choice of this metric is that the location of nodes and length of edges tend to reach a near static state in a 3D force-directed graph when the attractive and repulsive forces are being balanced. But after adding the geo-force, the nodes which are originally clustered together may be stretched far apart if their geolocations are distant from each other, resulting in an increase in the ELV.

We also introduce the mean locational offset (MLO) to measure the location accuracy in the visual presentation of the graph nodes. The metric $M_{MLO}$ is defined as follows:

\begin{align}
   M_{MLO} = \sum\limits_{v \in V}\frac{\left\|C(v) - G(v)\right\|}{n_V\cdot d_{GC}}
\end{align}

Where $V$ is the set of nodes in the graph, $n_V$ is the number of nodes, $C(v)$ indicates coordinate vector of a node in the final graph layout. $G(v)$ is the coordinate vector of the projected geolocation of a node. The offset is computed as the horizontal distance that a node moves in the 2D map plane. The distance between the North Pole and the South Pole on the map, $d_{GC}$, measured in the viewport coordinate system, is used to normalize the MLO. When every node is placed right above its exact geolocation, $M_{MLO}=0$. 

\begin{table}
  \centering
  \caption{Quantitative measures for graph structure change of expert data under different force balancing parameter $K$}
  \label{tab1}
  \begin{tabular}{ccrr}
    \toprule
    Scenario & $K$ & $M_{ELV}$ & $M_{MLO}$ \\
    \midrule
    (a) & 0      & 0.0555 & 0.554 \\
    (b) & 5      & 0.0731 & 0.119 \\
    (c) & 10,000 & 0.0941 & 0.001 \\
    \bottomrule
  \end{tabular}
\end{table}

\begin{figure*}[h]
  \begin{subfigure}{\textwidth}
      \centering
      \includegraphics[width=0.96\textwidth]{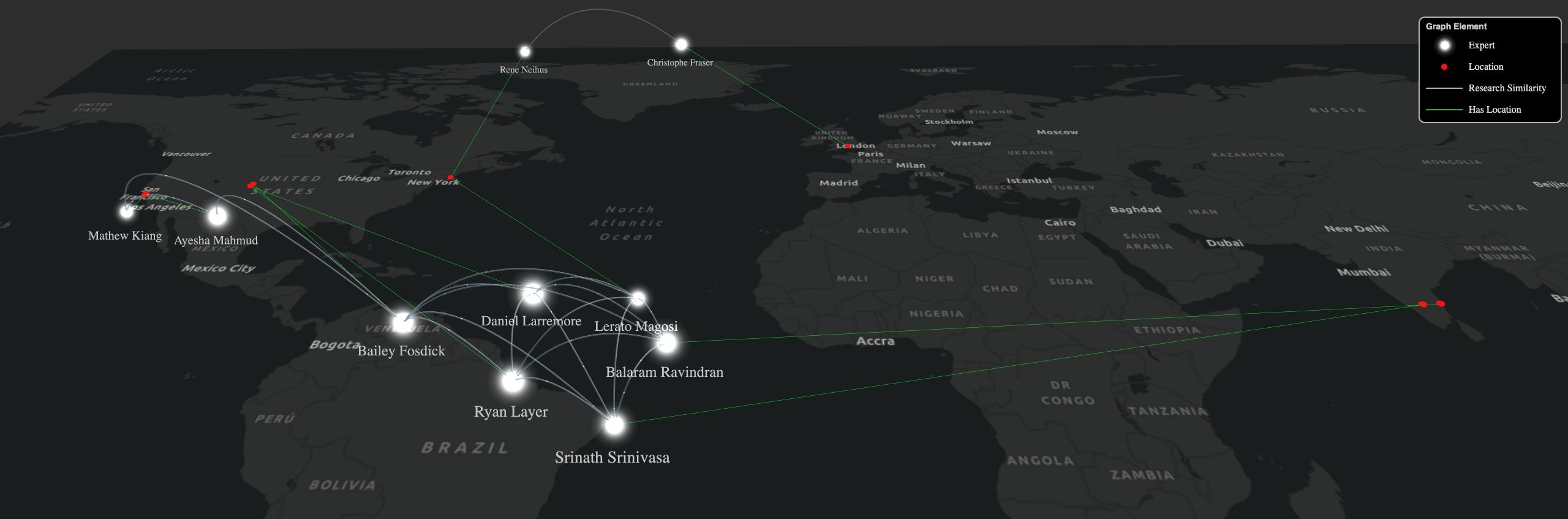}
      \caption{}
      \label{fig2:a}
  \end{subfigure}
  \begin{subfigure}{\textwidth}
      \centering
      \includegraphics[width=0.96\textwidth]{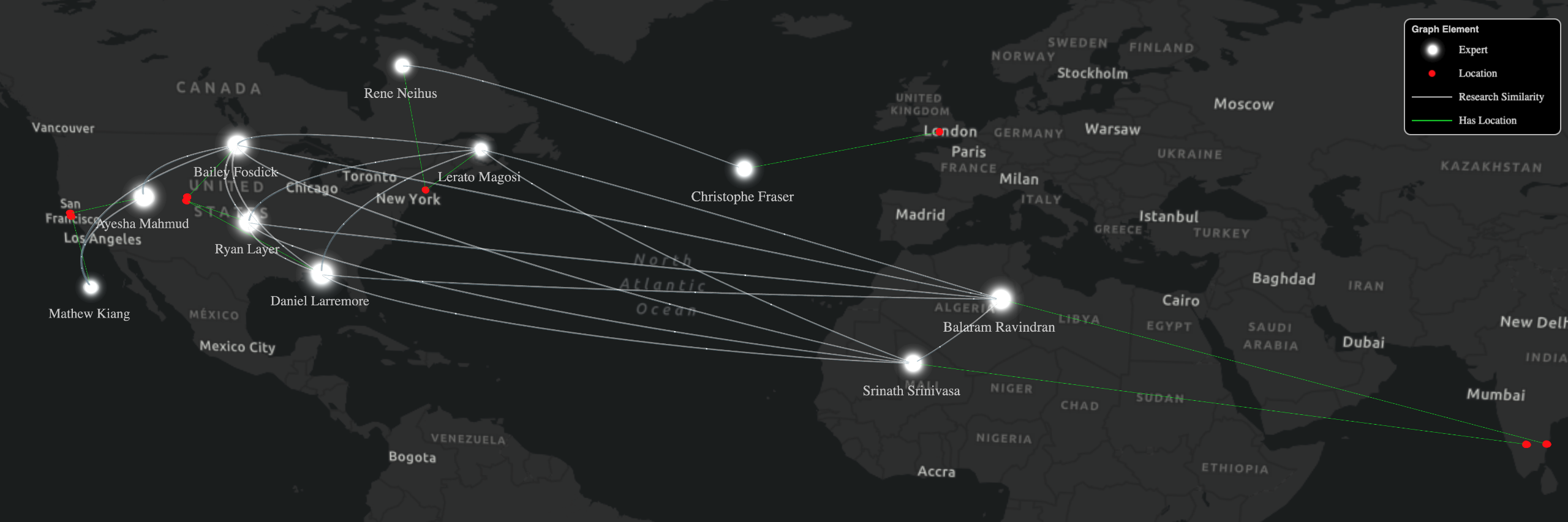}
      \caption{}
      \label{fig2:b}
  \end{subfigure}
  \begin{subfigure}{\textwidth}
      \centering
      \includegraphics[width=0.96\textwidth]{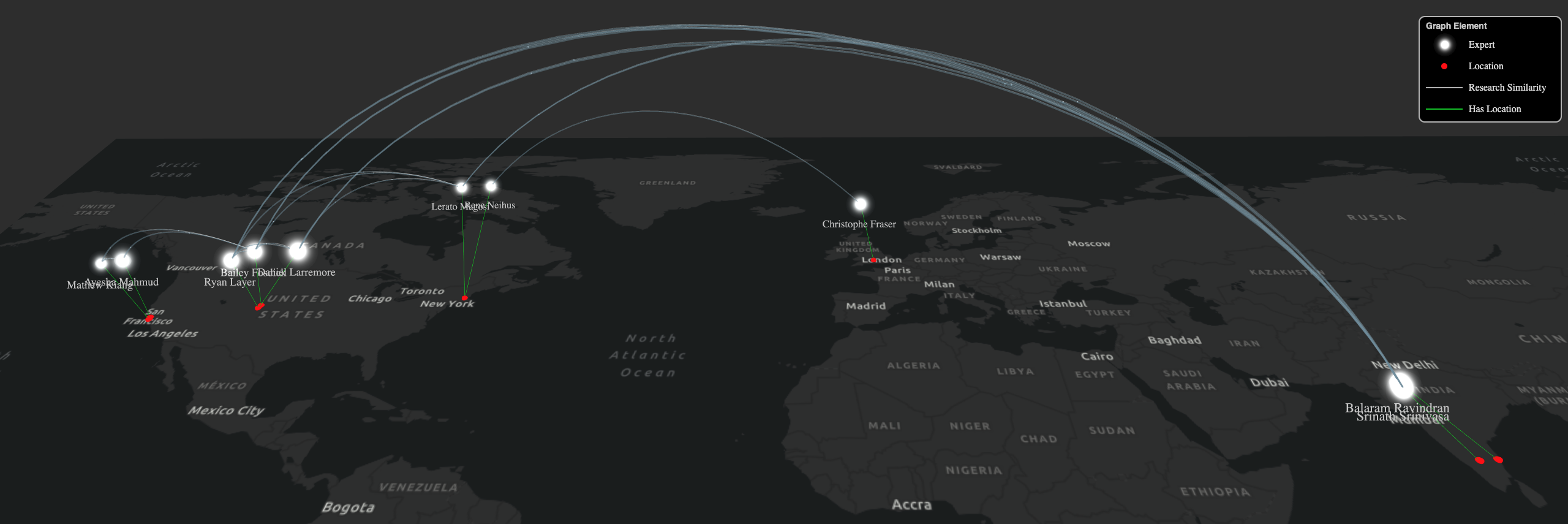}
      \caption{}
      \label{fig2:c}
  \end{subfigure}
  \caption{Graph layout at the force balancing parameter $K$. (a) $K=0$, (b) $K=5$, and (c) $K=10,000$}
  \label{fig2}
\end{figure*}

Table \ref{tab1} presents the values of $M_{ELV}$ and $M_{MLO}$ at different settings of the forcing balancing parameter $K$. The resultant graph layouts visualized by the GeoGraphViz are shown in Figure \ref{fig2}. To more clearly present the 3D graph through a 2D snapshot, less prominent nodes (nodes with fewer connections) are set to be invisible, but their interactions through the three forces are still considered. In practice, we recommend setting the parameter $K$ between 3 and 20. Graphs created with $K$ falling in this value range can better preserve location representation of each node, as well as important graph structures, such as local clusters. In this experiment, we use $K=5$ to present a balanced graph layout. Figure \ref{fig2:a} shows the scenario in which the geo-force is excluded when drawing the graph. In this case, $K=0$ and $M_{ELV}$ reaches its lower bound and $M_{MLO}$ reaches at its upper bound for this test dataset in which experts are distributed very broadly with a strong international perspective.  In the Figure \ref{fig2:a} graph, we can clearly observe a local cluster near the center of the graph. However, it is almost impossible to correctly infer the geographic distribution pattern of these experts, especially for the two experts from India, if we consider only the information in the graph layer (white nodes and white links). Here, because GeoGraphViz implements the new ``Graph Above the Map'' strategy and uses the green lines to connect the nodes with its geolocation on the map layer, the spatial context can still be captured. 

Figure \ref{fig2:c} presents a scenario when $K$ is set to be a very large number (Table \ref{tab1}), where the geo-force dominates the graph layout simulation. The graph nodes in Figure \ref{fig2:c} locate nearly exactly on top of their actual geolocation, showing a clear geographic distribution pattern. Hence, the $M_{MLO}$ is at a near-zero value. However, the graph structure is significantly varied with the local cluster presented in Figure \ref{fig2:a} almost completely gone. It is therefore very difficult to analyze the graph patterns through the graph's visual topological presentation. $M_{ELV}$ in this case reaches its upper bound due to the significant relocation of graph nodes. 

In comparison, Figure \ref{fig2:b} presents a more balanced view in location-aware graph visualization (with $K=5$). While the nodes are moving closer to reflect their geolocation, we can still observe a highly densely connected subgraph near the center of the view. This graph view can also better present the international perspective of the clustered nodes of experts with similar research expertise. Quantitatively, $M_{ELV}$ reduces from when $K=0$ and falls near the middle of its value range, meaning that nodes move but not dramatically. We also found that $M_{MLO}$ has reduced substantially compared to when $K=0$, meaning that the location offset is much reduced so the spatial context can be better presented. 

Observing value changes in the two metrics ($M_{ELV}$ and $M_{MLO}$) can also help reveal the joint effects of the attractive force, repulsive force, and the geo-force to the graph layout. Ideally, if the nodes of each densely connected subgraphs (clusters) in a graph are all located within a local geographical region, the algorithm is capable of generating a graph layout with both a clear graph pattern and spatial pattern. In such a scenario, $M_{ELV}$ and $M_{MLO}$ will remain low with varying $K$. However, when a moderate $K$ value (e.g., between 3 and 20) results in a large edge length variation (with $M_{ELV}$ at close to its upper bound), this means adding geo-force will significantly change the graph structure. For applications that focus on investigating the graph patterns, the use of geo-force is not recommended. Instead, the users can still rely on the GeoGraphViz’s unique ``Graph Above the Map'' feature (e.g., green lines connecting graph nodes with their geolocations) to understand the spatial distribution patterns. 

\subsection{Algorithm generalizability test} \label{subsec_exp_simul}
To further test the generalizability of the proposed algorithm, we created a simulated dataset with distinctive graph and geographical distribution patterns and more nodes (>200) than the real dataset used in Section 4.2. The graph can be considered as a large collaboration network of researchers with international perspectives. It can also be used to model semantic relationships for other types of data (e.g., publications, images, and commercial products), among which their similarity can be measured. 

\begin{figure*}[h]
  \begin{subfigure}{\textwidth}
      \centering
      \includegraphics[width=0.8\textwidth]{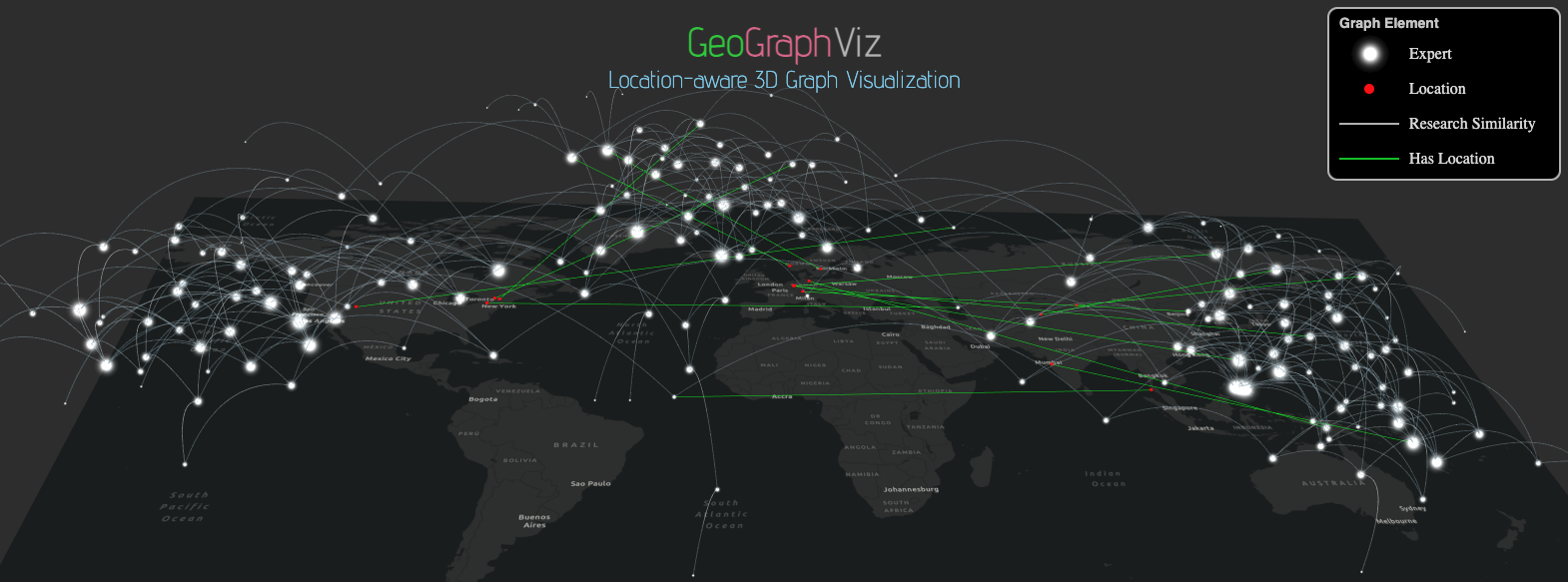}
      \caption{}
      \label{simres_k0}
  \end{subfigure}
  \begin{subfigure}{\textwidth}
      \centering
      \includegraphics[width=0.8\textwidth]{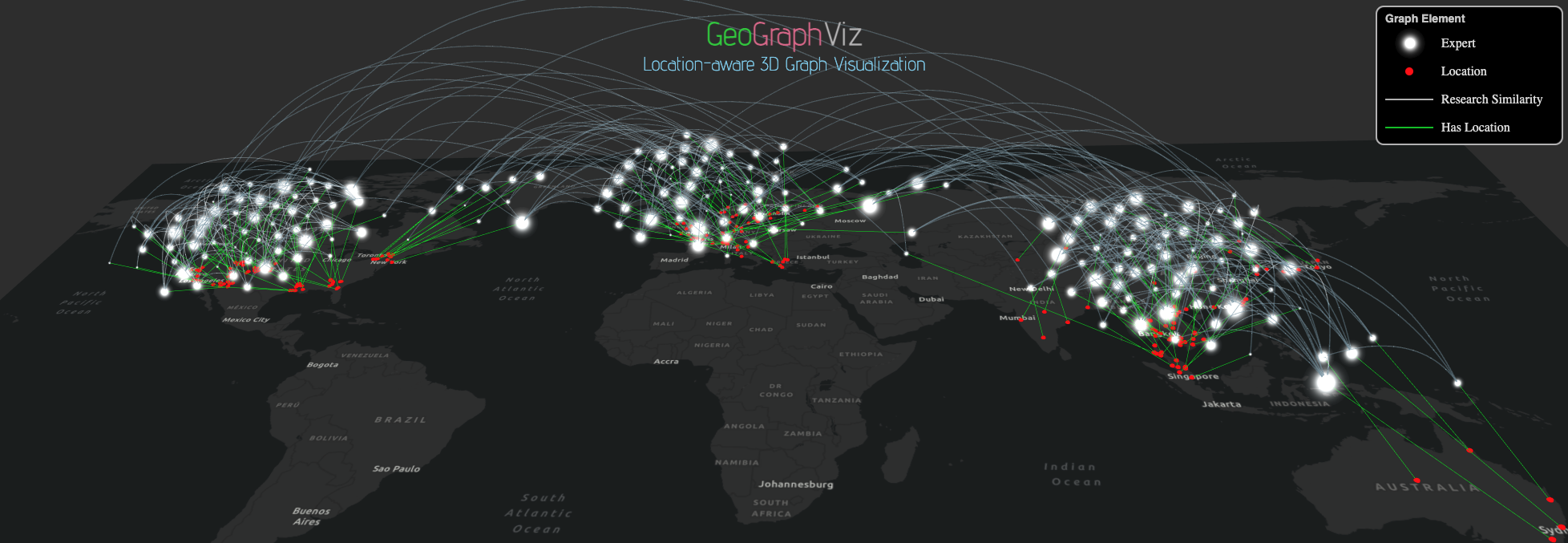}
      \caption{}
      \label{simres_k5}
  \end{subfigure}
  \begin{subfigure}{\textwidth}
      \centering
      \includegraphics[width=0.8\textwidth]{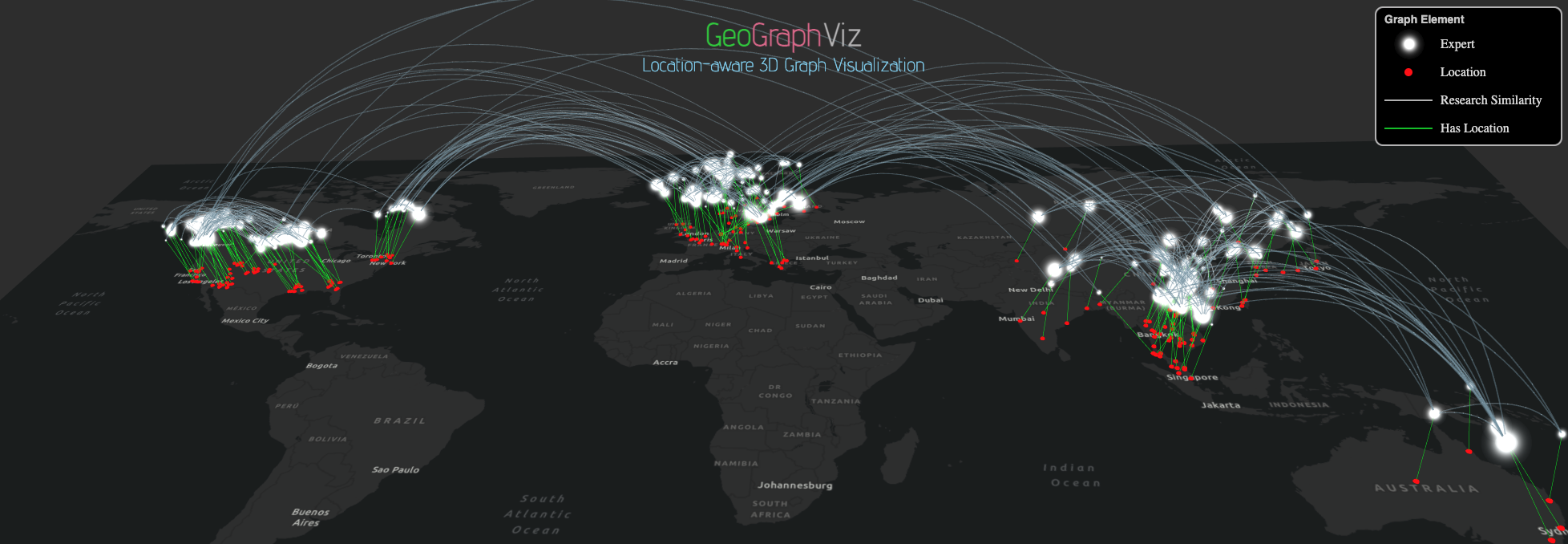}
      \caption{}
      \label{simres_k10k}
  \end{subfigure}
  \caption{Graph layout of simulated dataset at the force balancing parameter $K$. (a) $K=0$, (b) $K=5$, and (c) $K=10,000$}
  \label{simres}
\end{figure*}

This simulated dataset contains three major clusters. The number of nodes distributed to each cluster are about the same ($\sim$70). The similarity values among each pair of nodes were randomly generated, following normal distributions. The existence of a linkage between two nodes, whether they are in the same cluster or not, is based on a predefined probability. More connections (edges) with stronger strengths (higher similarity values) will be assigned to nodes within the same cluster, and the number of edges and similarity values are lower for between-cluster nodes. The white nodes and white links in Figure \ref{simres_k0} illustrate the clusters. This figure also shows when no geographical force is applied to each node. Hence, the clusters are solely semantic. To assess the effect of location awareness in graph visualization, each node in the simulated dataset was assigned with a geo-location. In general, nodes in each of the three clusters are geographically distributed in the United States (US), Europe, and Asia, respectively. From west to east, let us call the three main clusters observed (in Figure \ref{simres_k0}) the U.S., the European, and the Asian cluster. However, outliers also exist. The green linkages in Figure \ref{simres_k0} show nodes which "travel" a long distance. For instance, several nodes whose geolocations are in the U.S. (the New York region) belong semantically to the European cluster. When there are no geographical constraints, these nodes will be placed near nodes to which they are akin. Similar cases also include (1) a few nodes geographically located in Germany but semantically belong to the Asian cluster; and (2) three nodes in Norway, Sweden, and Thailand moved westward to be close to the clusters to which they belong. Nodes with significant geographical movements are highlighted in greenlinks in Figure \ref{simres_k0}. 

\begin{table}
  \centering
  \caption{Quantitative measures for graph structure change of simulated data under different force balancing parameter $K$}
  \label{tab2}
  \begin{tabular}{ccrr}
    \toprule
    Scenario & $K$ & $M_{ELV}$ & $M_{MLO}$ \\
    \midrule
    (a) & 0      & 0.0241 & 0.315 \\
    (b) & 5      & 0.0445 & 0.0576 \\
    (c) & 10,000 & 0.0519 & 5.68e-5 \\
    \bottomrule
  \end{tabular}
\end{table}

The visualization results of the simulated graph under different force balancing parameter $K$ are shown in Figure \ref{simres}. While Figure \ref{simres_k0} demonstrates the visualized graph without a geo-force, Figure\ref{simres_k10k} shows when a very strong geo-force is applied ($K$=10,000) and has become the dominant force. In such a case, all nodes are placed very close to their geo-locations, so it is easy to observe the geographical properties of the graph nodes, but the graph's semantic structure can hardly be inspected. A visualization result that balances between the two forces is shown in Figure \ref{simres_k5}, which presents a clear spatial distribution as well as the semantic clustering pattern. In particular, the nodes from New York and Germany are dragged close to their actual clusters, but their geo-locations are still recognizable in the given parameter settings. 

The statistics of the above three configurations, measured by $M_{ELV}$ and $M_{MLO}$, are listed in Table \ref{tab2}. It demonstrates a similar pattern as that in Table \ref{tab1}. When $K$ increases from 0 to 5, the node's semantic distance ($M_{ELV}$) becomes twice as great, but the geographical distance $M_{MLO}$ drops significantly, reaching a balanced view between graph and spatial patterns.

\subsection{Algorithm Efficiency Test} \label{subsec_exp_effi}

To assess system efficiency in visualizing graphs with an increasing size, we simulated graph data with two types of patterns. In Type I graphs, the total number of graph edges are in proportion to that of a complete, undirected graph (i.e., for any two nodes in the graph, there is an edge connecting them). The number of edges $N_e$ in such graphs can be represented as:

\begin{equation}
    N_e = p \cdot \frac{n \cdot (n-1)}{2} 
\end{equation}

Where $n$ is the number of nodes in the graph. It can be seen that $N_e$ is proportional to $n^2$ controlled by a parameter $p \in (0,1]$.

Type II graphs share the characteristic of the total number of edges, being proportional to the number of graph nodes. Hence, $N'_e$ can be represented as:

\begin{equation}
    N'_e = \frac{c \cdot n}{2} \;,\;\; (c \leq n-1)
\end{equation}

Where $n$ is the number of nodes in the graph. In graphs with such patterns, the total number of edges, $N'_e$ is proportional to the number of the nodes. This proportion is controlled by a parameter $c \in [0,n-1]$. This strategy emulates real-world scenarios, such as social networks wherein each person maintains a certain number of (social) connections on average, regardless of the size of the network/graph.

\begin{figure*}
  \centering
  \includegraphics[width=0.66\textwidth]{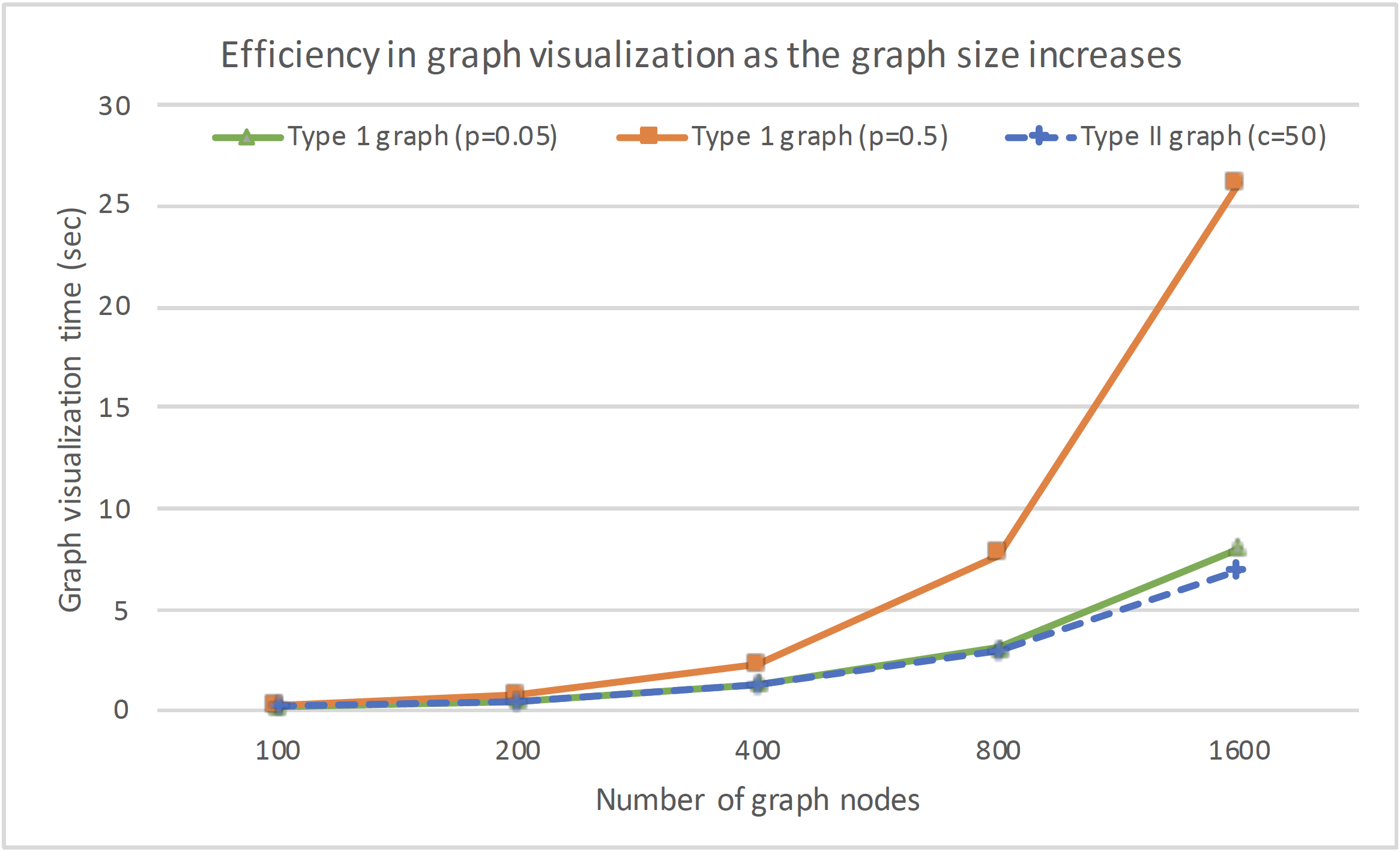}
  \caption{Efficiency of graph visualization as graph size increases}
  \label{exp_effi}
\end{figure*}

We further generated graphs with different sizes (with $n$ = 100, 200, 400, 800, and 1600) and different graph density parameters ($p$=0.05, $p$=0.5, and $c$=50) following the two graph types. The graph visualization time counted from loading the graph data, and simulating the positions of nodes in the graph, to finishing rendering it in the web browser, as reported in Figure \ref{exp_effi}. 

The X axis shows the number of nodes in the graph. The Y axis provides the graph visualization time (in sec). In Type I graphs, the number of edges is proportional to the square of number of nodes. The larger the parameter $p$, the denser the graph. In Type II graphs, the number of edges is proportional to the number of nodes. The larger the parameter $c$, the denser the graph. 
The dashed blue line demonstrates the result of a graph simulation using the second strategy (Type II graph) with $c$=50. The solid green line with triangular markers indicates the visualization time for Type I graphs with $p$=0.05. The solid orange line with square markers shows the results of Type I graphs with $p$=0.5. It also presents an edge case (rare in real-world scenarios) where the graphs are very densely connected (the number of edges equal to half of that in a undirected, complete graph). The green and blue lines show similar results. As the number of nodes increases, there is a linear relationship between visualization efficiency and the number of graph nodes. The orange line, in comparison, shows almost a quadratic growth per number of nodes. This result reflects the time complexity of $O(n^2)$ indicated in the algorithm presented in Section \ref{sec_alg}.  

As seen in Figure \ref{exp_effi}, the performance of the GeoGraphViz algorithm in terms of time efficiency is good and can serve near real-time visualization purposes when the number of graph nodes is below 800. As shown in both the dashed blue and solid green lines, the visualization time will be less than 3 sec. For the orange line, the visualization can be completed within 7 sec. There are some lags but considering the density of the graph, this visualization time is acceptable. When the number of graph nodes reaches a very high number, such as 1600, the lags become longer. In such cases, further optimization is required to improve both the visualization efficiency and the visual effect, as it will be difficult to examine the pattern of a very large graph with dense connections.

\section{GeoGraphViz User Interface}

\begin{figure*}
  \centering
  \includegraphics[width=0.96\textwidth]{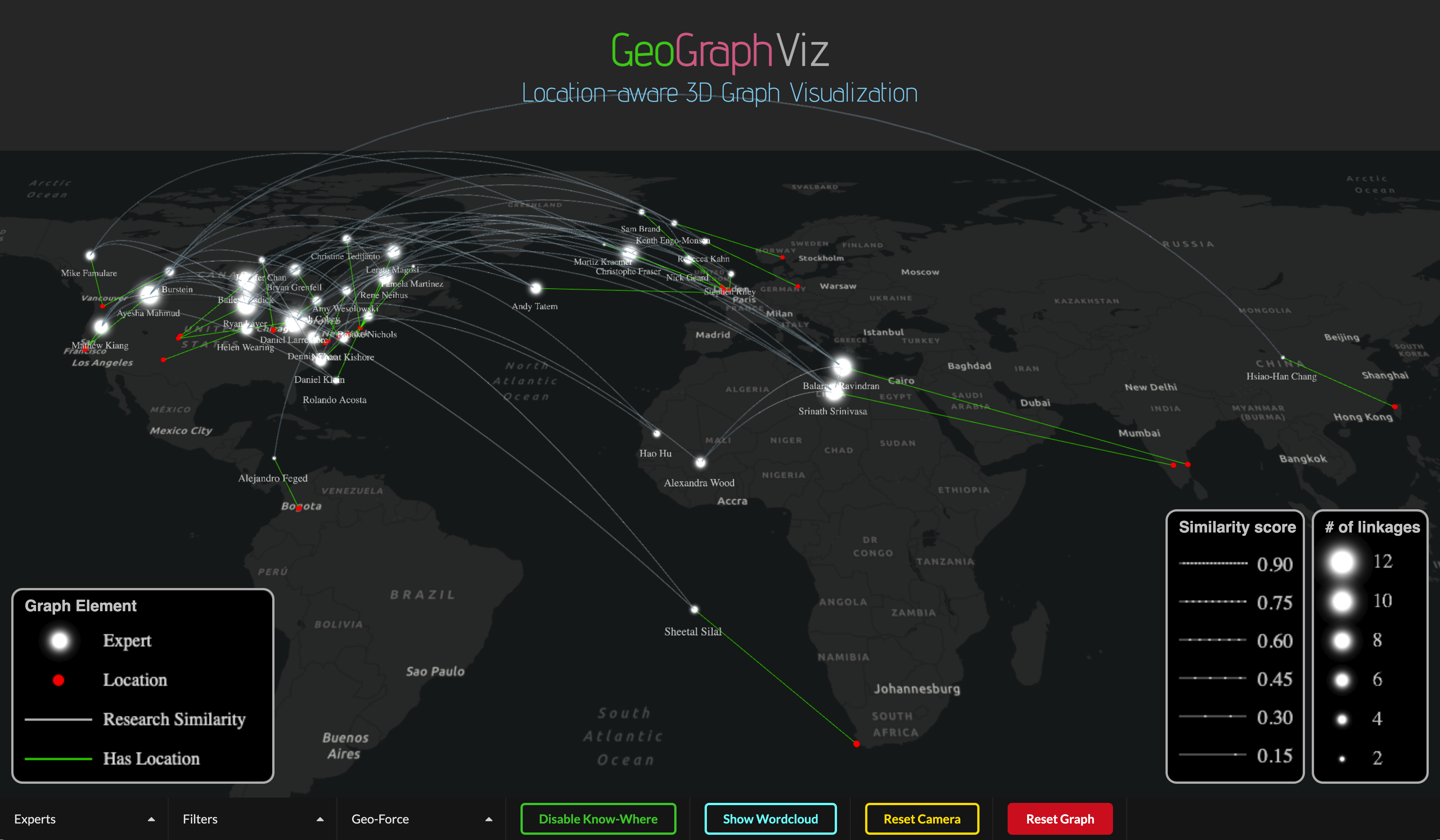}
  \caption{The GeoGraphViz user interface for location-aware knowledge graph visualization}
  \label{gui-overview}
\end{figure*}

Figure \ref{gui-overview} demonstrates the interface for our graph visualization tool (with expert data). As seen, there are two layers shown in the interface: (1) the 3D graph layer showing connections of disease experts (white lines connecting white nodes); and (2) a map layer connecting experts to their locations (i.e., geolocations of their affiliations) by using green lines connecting white nodes to the red dots. These two layers together could demonstrate the clusters of disease experts who share similar research interests from the graph perspective. They can also reveal the spatial patterns of the potential collaboration network, be it local, regional, or international. Several supporting functions are developed to allow (1) clicking on a graph node to view the profile information of an expert (see an example in Figure \ref{fig1}), (2) filtering the graph to present subgraphs with different degrees of node connections (Figure \ref{fig2}), (3) turning on and off the ``KnowWhere'' feature to display purely a graph view or the “Graph Above the Map” view, and (4) a word cloud view showing collective expertise of the experts in a selected subgraph. Many graph features (e.g., connections among the graph nodes and connections across graph and geolocations) can be set to invisible to allow examination of different facets of the graph. The graph can also be rotated, panned, and zoomed to improve user experience. 

\section{Conclusion}

This paper presents GeoGraphViz, a location-aware knowledge graph visualization tool to allow end users, both domain experts and the general public, to interactively explore graph data, to identify hidden patterns, and most importantly, to support discovery of new (geospatial) knowledge from massive cross-domain datasets. GeoGraphViz is empowered by a novel geographically constrained, 3D force-directed, graph visualization algorithm and it addresses the limitations of general graph visualization methods and map-based visualizations and is capable of (1) intuitively visualizing location properties of the graph nodes, (2) performing multi-dimensional graph visualization for heuristic knowledge search, and (3) achieving dynamic, interactive, and context-aware graph visualization. Currently, our GeoGraphViz tool is being used in combination with one of the largest geospatial knowledge graphs, the KnowWhereGraph, to empower environmental intelligence applications in the areas of disaster response and humanitarian aid, climate change, and agricultural production to food supply chain management.

Leveraging a three-dimensional dynamic graph visualization and a strategy of ``Graph Above the Map,'' our GeoGraphViz tool has the unique strength of ``knowing where.'' We described the mathematical formulation of the algorithm in Section \ref{sec_alg}, as well as conducted a series of experiments in Section \ref{sec_exp} to verify the efficiency and generalizability of the algorithm in visualizing graphs with different patterns. In Section \ref{subsec_exp_simul}, we simulated graph data in which the geographical and semantic distribution patterns share some similarities (i.e., three main semantic clusters reside mainly in three regions: US, Europe, and Asia), but there are exceptions – while a certain number of nodes reside in one main geographical region, they belong semantically to another semantic cluster(s). In such cases, as the geo-force is gradually applied, we can observe a change in the graph pattern where a node can move a very long distance (examples shown in Figure \ref{simres_k0}) to where a balanced view is achieved between the proximity of this node to its geographical location and the proximity of this node to the other nodes semantically similar to it (examples shown in Figure \ref{simres_k5}), and to where an edge case is reached, where every node is located right above its location and the semantic pattern be barely observable (Figure \ref{simres_k10k}).  

As seen in Figure \ref{simres_k5}, our algorithm is capable of stretching out each cluster (EU, Asia, and US) over the map real-estate and highlighting nodes that have a high level of connectivity to another cluster. This way, the mixed semantic and geographical patterns can be clearly presented. In fact, this graph layout and every other layout are simulated based on the overall attractive and repulsive force among all nodes, and the layout becomes stable when a global force equilibrium is achieved. However, when a dataset contains different similarity patterns, the graph will show different patterns. For instance, if the EU cluster receives an equal amount of total attractive force from the two edge clusters (US and Asia), the nodes in the EU cluster will not move significantly towards either. Instead, the Asia and US nodes will likely move toward the EU nodes, and the distance of movement will depend on the overall forces they receive.

In an extreme case, when the geographical clustering and semantic clustering are completely different, meaning that nodes in a semantic cluster do not have any geographic proximity, and can come from any place on the Earth's surface, then mixing the two forces (geographical and semantic) will not present a clear visual effect. In such cases, we recommend applying either the semantic or the geographical force in the visualization. Our GeoGraphViz tool provides functions to allow for such configurations. As known, visualizing multiple dimensions of information simultaneously helps us gain a more comprehensive view of a dataset, but it can become more challenging from a visualization standpoint. Our research aims to solve this problem. Although not all cases are suitable for being visualized using mixed force, we were able to offer a new way to present multi-dimensional information in the same visual space, and developed strategies to separately visualize the two patterns (geographical and semantic) in complex cases. 

In Section \ref{subsec_exp_effi}, we also conducted experiments to examine the efficiency of the graph visualization algorithm. We observed that when the graph size becomes very large (with the number of nodes above 800), the visualization lag time will become long, and this will affect the user’s experience. With a large graph, the nodes will likely be cluttered together. To address this, further optimization strategies will be needed to accelerate rendering efficiency and to improve the visualization effectiveness. Node clustering strategies, such as geometry-based edge bundling \cite{cui2008geometry} and multi-level edge agglomeration \cite{gansner2011multilevel}, are to be exploited and integrated. In addition, converting from CPU-based visualization to GPU (Graphics Processing Unit) visualization \cite{li2017polarglobe}\cite{wang2019capturing} could also be an effective means to help the algorithm reduce graph layout computation time and achieve desired real-time rendering performance.  

Our research has broader implications for advances in GIScience. First, the ``Graph on the Map'' solution offers a new way of presenting multi-dimensional information and has the potential to spark the development of more innovative geovisualization methods. Second, this new visualization strategy allows for a more visual, intuitive, and interactive way to explore geospatially enabled graph data, contributing to a better understanding of complex patterns and relationships, especially in a geographical context. Third, the visualization tool helps to provide a comprehensive and nuanced view of linked data, making it easier to communicate scientific information with stakeholders to make informed decisions. Finally, this work emphasizes the importance of ``location-awareness'' in knowledge graph research, which can further foster cross-domain knowledge exchange between the Semantic Web community and the GIS community.

In the future, we will continue to refine the visualization algorithm to enhance its spatial-semantic visual presentation and develop more search and filtering functions to better support graph knowledge exploration and intelligent scientific question answering. The data and code will be openly shared to benefit the broader research community.

\section{Data and code}
The data and code used for the GeoGraphVis system and the experiments can be accessed through: https://github.com/ASUcicilab/GeoGraphViz.

\section{Acknowledgements} 
This work is supported in part by the National Science Foundation under awards number 2033521, 1853864, and 2120943. Any opinions and findings expressed in this material are those of the authors and do not necessarily reflect the views of the National Science Foundation.

\bibliography{geographviz}

\end{document}